# Positive results from UK single gene PCR testing for SARS-COV-2 may be inconclusive, negative or detecting past infections

Prof. Martin Neil,

School of Electronic Engineering and Computer Science,

Queen Mary, University of London

18 March 2021 (version 7)

## Abstract

The UK Office for National Statistics (ONS) publish a regular infection survey that reports data on positive RT-PCR test results for SARS-COV-2 virus. This survey reports that a large proportion of positive test results may be based on the detection of a single target gene rather than on two or more target genes as required in the manufacturer instructions for use, and by the WHO in their emergency use assessment. Without diagnostic validation, for both the original virus and any variants, it is not clear what can be concluded from a positive test resulting from a single target gene call, especially if there was no confirmatory testing. Given this, many of the reported positive results may be inconclusive, negative or from people who suffered past infection for SARS-COV-2.

## Background

The efficacy of mass population testing for SARS-COV-2 virus is critically dependent on the reliability of the test applied, whether it be a RT-PCR or lateral flow test. Given that many RT-PCR tests do not actually target all the genes necessary to reliably detect SARS-COV-2, the results of mass testing using RT-PCR need to be revisited and reanalysed.

The ONS publish a regular infection survey [1], [20] that includes data from two UK lighthouse laboratories, based in Glasgow and Milton Keynes, where both use the same RT-PCR test kit, to detect the SARS-COV-2 virus. This survey includes data on the cycle threshold (Ct) used to detect positive samples, the percentage of positive test results arising from using RT-PCR, and the combinations of the SARS-COV-2 virus target genes tested that gave rise to positives between 21 September 2020 and 1 March 2021 across the whole of the UK.

The kit used by the Glasgow and Milton Keynes lighthouse laboratories is the ThermoFisher TaqPath RT-PCR[1] which tests for the presence of three target genes from SARS-COV-2[2] [11]. Despite Corman et al [2] originating the use of PCR testing for SARS-COV-2 genes[3] there is no agreed international standard for SARS-COV-2 testing. Instead, the World Health Organisation (WHO) leaves it up to the manufacturer to determine what genes to use and instructs end users to adhere to the manufacturer instructions for use (IFU). As a result of this we now have an opaque plethora of commercially available testing kits, that can be applied using a variety of test criteria. Other UK laboratories use different testing kit, and test for different genes.

[1] The full name for ThermoFisher TaqPath kit is TaqPath COVID-19 CE-IVD RT-PCR.

[2] N, S and ORF1ab genes

[3] Corman et al recommended the E, N and RdRp genes

The WHO's emergency use assessment (EUA) for the ThermoFisher TaqPath kit [3] includes the instruction manual and contained therein is an interpretation algorithm describing an unequivocal requirement that two or more target genes be detected before a positive result can be declared. This is shown in Table 1. The latest revision of ThermoFisher's instruction manual contains the same algorithm [21].

**Table 6 Result interpretation for patient samples**

| ORF1ab | N gene | S gene | MS2 | Status | Result | Action |
|---|---|---|---|---|---|---|
| NEG | NEG | NEG | NEG | INVALID | NA | Repeat test. If the repeat result remains invalid, consider collecting a new specimen. |
| NEG | NEG | NEG | POS | VALID | SARS-CoV-2 Not Detected | Report results to healthcare provider. Consider testing for other viruses. |
| Only one SARS-CoV-2 target = POS | | | POS or NEG | VALID | SARS-CoV-2 Inconclusive[1] | Repeat test. If the repeat result remains inconclusive, additional confirmation testing should be conducted if clinically indicated. |
| Two or more SARS-CoV-2 targets = POS | | | POS or NEG | VALID | Positive SARS-CoV-2 | Report results to healthcare provider and appropriate public health authorities, as applicable. |

[1] Samples with a result of SARS-CoV-2 Inconclusive shall be retested one time. Retesting must be performed from the biologial sample originally collected from the patient.

**Table 1: Screenshot of results interpretation ThermoFisher TaqPath IFU on page 60 of [3] (their Table 6)**

The WHO have been so concerned about correct use of RT-PCR kit that on 20 January 2021 they issued a notice for PCR users imploring them to review manufacturer IFUs carefully and adhere to them fully [4].

## Increasing proportion of single gene target "calls"

The ONS's report [1] lists SARS-COV-2 positive results for valid two and three target gene combinations[4] and does the same in [20], for samples processed by the Glasgow and Milton Keynes lighthouse laboratories. However, it also lists single gene detections as positive results[5] (See tables 6a and 6b). This use of single gene "calls" suggests that these lighthouse laboratories may have breached WHO emergency use assessment (EUA) and potentially violated the manufacturer instructions for use (IFU). According to the WHO, such single gene calls should be classified as inconclusive test results. However, Section 10 of this ONS Covid-19 Infection survey report [5] on the 8 January 2021 stated that one gene is sufficient for a positive result (emphasis mine):

> "Swabs are tested for three genes present in the coronavirus: N protein, S protein and ORF1ab. Each swab can have any one, any two or all three genes detected. Positives are those where **one** or more of these genes is detected in the swab ….."

Over the period reported the maximum weekly percentage of *positives on a single gene* is 38% for the whole of the UK for the week of 1 February. The overall UK average was 23%. The maximum percentage reported is 65%, in East England in the week beginning 5 October. In Wales it was 50%, in Northern Ireland it is 55% and in Scotland it was 56%. The full data including averages and maxima/minima are given in Table 2.

[4] N+S+ORF, ORF+S, N+S and N+ORF gene combinations

[5] N alone, ORF alone (note that the S gene is included in the ONS analysis but is never counted as a positive if it is detected in isolation)

Figures 1 and 2 show the percentage of weekly single gene positives across the UK nations and English regions. There has been a significant increase in the percentage of single gene positives since the end of 2020, rising from January, and here the rise is steady across all English regions and UK nations.

| **Percentage of positive cases with single target gene** | UK | England | Wales | NI | Scotland | North East | North West | Yorkshire and Humber | East Midlands | West Midlands | East of England | London | South East | South West |
|---|---|---|---|---|---|---|---|---|---|---|---|---|---|---|
| 21 September 2020 | 17 | 17 | 0 | 33 | 0 | 6 | 14 | 22 | 40 | 9 | 18 | 9 | 40 | 58 |
| 28 September 2020 | 13 | 13 | 0 | 0 | 0 | 4 | 13 | 10 | 26 | 9 | 25 | 13 | 24 | 16 |
| 5 October 2020 | 19 | 19 | 14 | 6 | 16 | 8 | 9 | 9 | 11 | 16 | 65 | 32 | 58 | 34 |
| 12 October 2020 | 15 | 15 | 16 | 14 | 21 | 16 | 14 | 13 | 16 | 15 | 19 | 19 | 19 | 18 |
| 19 October 2020 | 19 | 19 | 34 | 24 | 13 | 28 | 12 | 17 | 25 | 18 | 28 | 28 | 19 | 21 |
| 26 October 2020 | 13 | 15 | 4 | 25 | 5 | 11 | 11 | 12 | 17 | 14 | 20 | 16 | 21 | 15 |
| 2 November 2020 | 16 | 16 | 18 | 10 | 23 | 18 | 13 | 13 | 17 | 14 | 23 | 25 | 17 | 13 |
| 9 November 2020 | 21 | 20 | 25 | 40 | 29 | 14 | 19 | 19 | 20 | 25 | 23 | 18 | 21 | 24 |
| 16 November 2020 | 17 | 17 | 10 | 6 | 22 | 11 | 25 | 20 | 18 | 19 | 13 | 13 | 13 | 16 |
| 23 November 2020 | 24 | 23 | 44 | 7 | 34 | 30 | 28 | 26 | 35 | 21 | 22 | 9 | 16 | 22 |
| 30 November 2020 | 29 | 29 | 25 | 18 | 32 | 46 | 40 | 49 | 28 | 24 | 18 | 17 | 29 | 22 |
| 7 December 2020 | 27 | 27 | 21 | 13 | 30 | 47 | 38 | 44 | 25 | 37 | 10 | 13 | 29 | 42 |
| 14 December 2020 | 15 | 15 | 1 | 7 | 29 | 20 | 20 | 23 | 19 | 24 | 10 | 9 | 11 | 31 |
| 21 December 2020 | 13 | 13 | 12 | 0 | 32 | 15 | 25 | 16 | 14 | 18 | 12 | 7 | 9 | 40 |
| 28 December 2020 | 20 | 20 | 19 | 10 | 31 | 29 | 22 | 29 | 19 | 16 | 17 | 15 | 16 | 36 |
| 4 January 2021 | 17 | 16 | 22 | 16 | 35 | 9 | 18 | 22 | 19 | 9 | 11 | 16 | 16 | 22 |
| 11 January 2021 | 29 | 28 | 48 | 23 | 38 | 28 | 29 | 43 | 31 | 18 | 36 | 25 | 28 | 25 |
| 18 January 2021 | 33 | 32 | 42 | 38 | 48 | 41 | 25 | 34 | 32 | 30 | 32 | 31 | 32 | 42 |
| 25 January 2021 | 35 | 33 | 50 | 42 | 35 | 25 | 28 | 28 | 34 | 31 | 37 | 36 | 38 | 39 |
| 1 February 2021 | 38 | 39 | 39 | 49 | 30 | 22 | 28 | 27 | 31 | 30 | 49 | 43 | 47 | 37 |
| 8 February 2021 | 35 | 36 | 16 | 55 | 34 | 32 | 31 | 28 | 35 | 28 | 41 | 40 | 38 | 43 |
| 15 February 2021 | 33 | 33 | 38 | 24 | 39 | 32 | 29 | 33 | 26 | 40 | 32 | 37 | 31 | 26 |
| 22 February 2021 | 31 | 32 | 33 | 9 | 36 | 15 | 30 | 30 | 38 | 35 | 31 | 34 | 32 | 25 |
| 1 March 2021 | 33 | 30 | 44 | 36 | 56 | 50 | 15 | 23 | 38 | 29 | 37 | 40 | 32 | 21 |
| Average | 23 | 23 | 24 | 21 | 28 | 23 | 22 | 25 | 26 | 22 | 26 | 23 | 27 | 29 |
| Min | 13 | 13 | 0 | 0 | 0 | 4 | 9 | 9 | 11 | 9 | 10 | 7 | 9 | 13 |
| Max | **38** | **39** | **50** | **55** | **56** | **50** | **40** | **49** | **40** | **40** | **65** | **43** | **58** | **58** |

**Table 2: Percentage of weekly single gene positives from 21 September 2020 to 1 March 2021, including averages and maxima/minima**

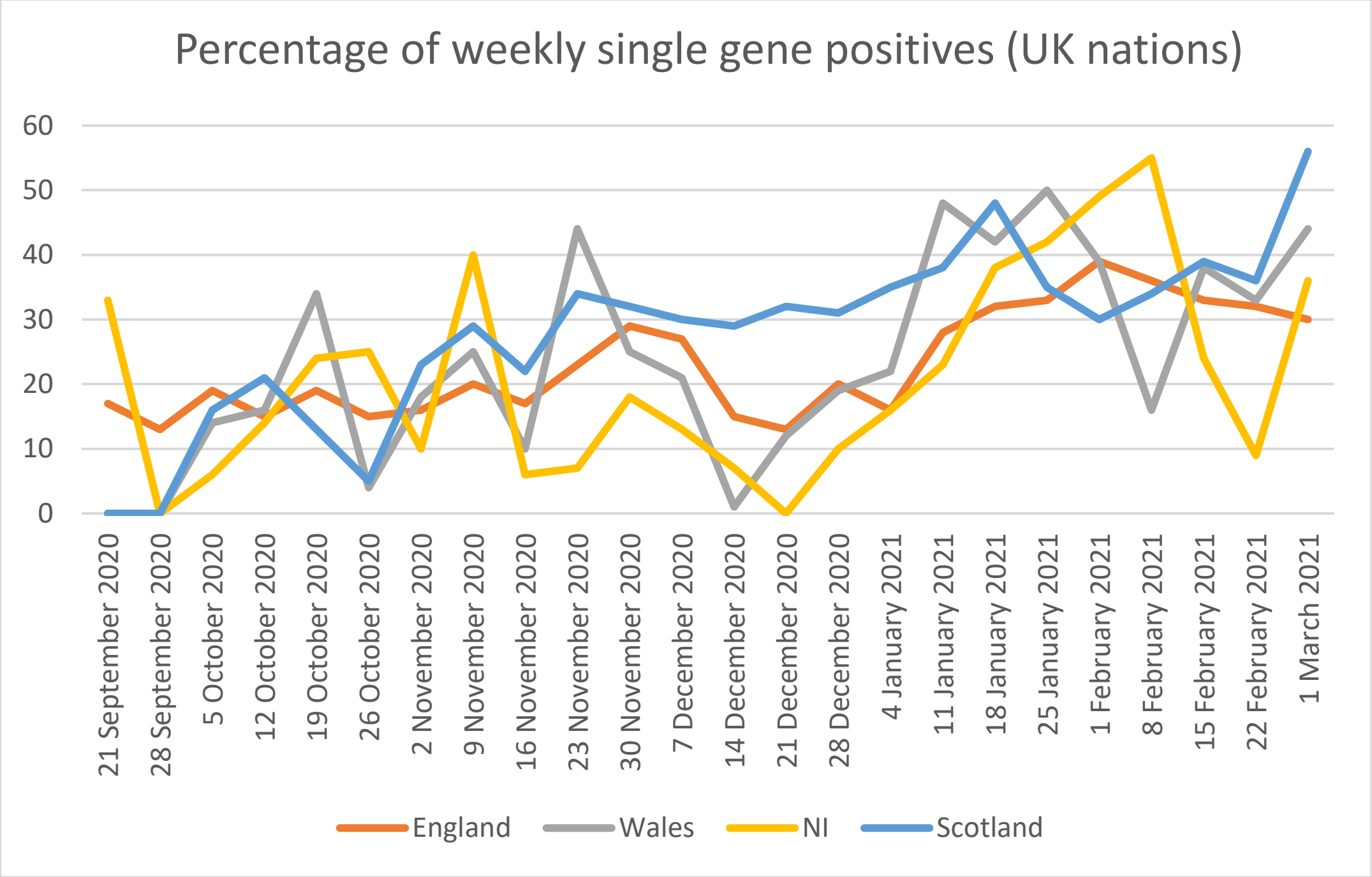

**Figure 1: Percentage of weekly single gene positives from 21 September 2020 to 25 January 2021 (UK nations)**

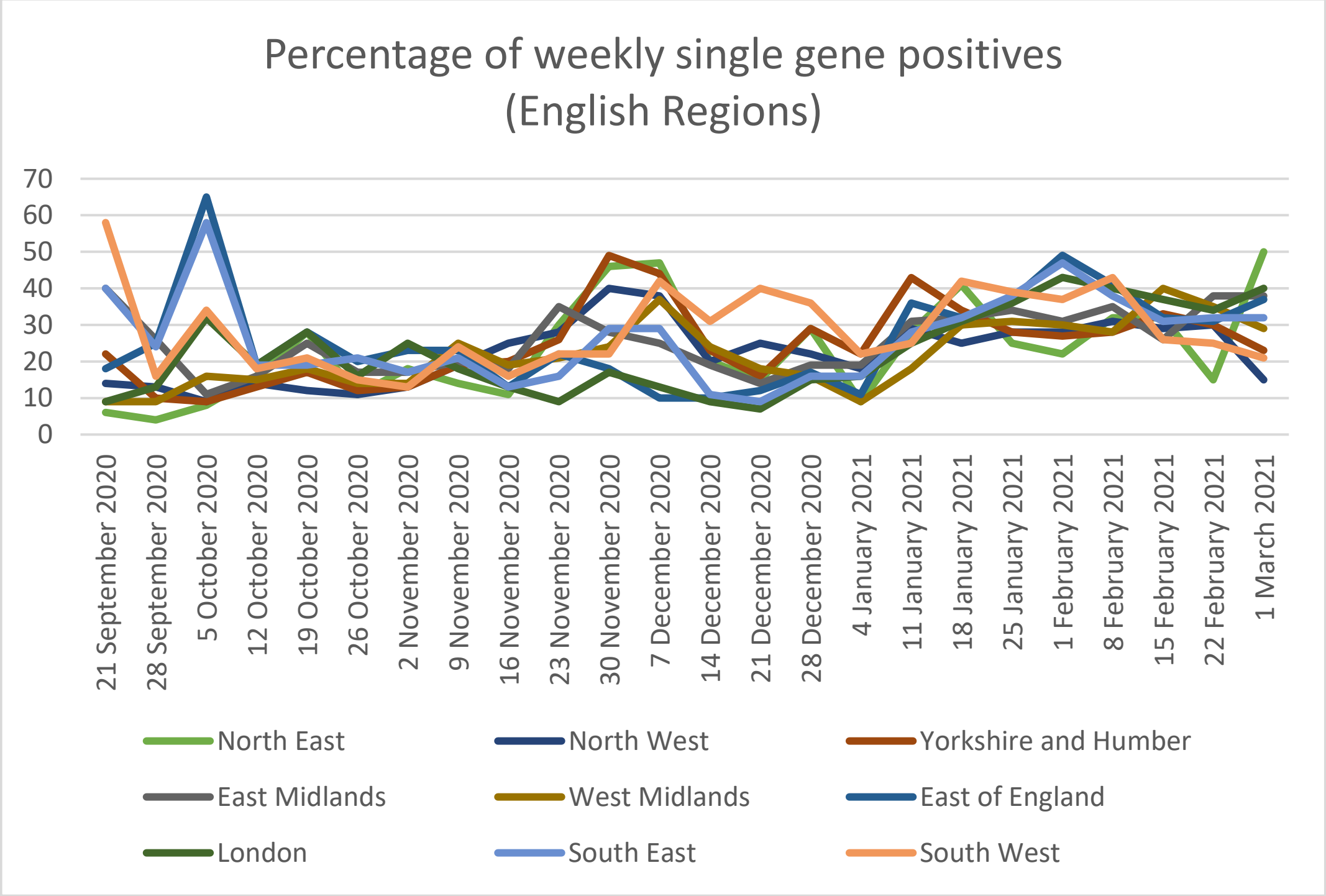

**Figure 2: Percentage of weekly single gene positives from 21 September 2020 to 25 January 2021 (English regions)**

Professor Alan McNally, Director of the University of Birmingham Turnkey laboratory, who helped set up the Milton Keynes lighthouse laboratory, contradicted what was stated in the ONS report in a Guardian newspaper article about the new variant. He reported that all lighthouse laboratories operated a policy that adhered to the manufacturer instructions for use:

requiring two-or-more genes for positive detection [6] (this policy is also documented in [22], which defines the standard operating procedure reported in [7]).

In correspondence with Mr Nicholas Lewis about single gene testing, in February 2021, the ONS confirmed that they do indeed call single gene targets as positives in their Covid-19 Infection Survey and also confirmed that the samples are processed by UK lighthouse laboratories [8], [9].

As early as April 2020, the UK lighthouse laboratories were testing for single genes and discounted the S gene as early as mid-May [10], months before the discovery of the new variant B1.1.7 (emphasis mine):

> "Swabs were analysed at the UK's national Lighthouse Laboratories at Milton Keynes (National Biocentre) (from 26 April) and Glasgow (from 16 August) …., with swabs from specific regions sent consistently to one laboratory. RT-PCR for three SARS-CoV-2 genes (N protein, S protein and ORF1ab) ..... Samples are called positive in the presence of at least **single** N gene and/or ORF1ab but may be accompanied with S gene (1, 2 or 3 gene positives). S gene is not considered a reliable single gene positive (as of mid-May 2020)."

Indeed, in Table 1 of [10] 18% of tests were positive on one gene only and it was concluded, in Table 2 of [10] that, for people with single gene positives, when Ct > 34, none had symptoms and for people with Ct < 34 only 33% had symptoms.

Furthermore in a Public Health England report on variants [11], published January 8th 2021, it states the goal of using one gene was explicitly to approximate the growth of the new B1.1.7 variant (emphasis mine):

> "There has recently been an increase in the percentage of positive cases where **only** the ORF1ab- and N-genes were found and a decrease in the percentage of cases with all three genes. We can use this information to **approximate** the growth of the new variant."

## Quality control and cross reactivity

Quality control problems have already been reported in UK laboratories [12, 13, 14] and concerns have been expressed about the potential for false positives arising consequently. Recent suspicion focused on problems potentially caused by exceeding acceptable Ct thresholds, suggesting no, or past, infection. However, this new ONS data shows there may be an additional potentially dominant source of false positives, at least within the period covered by the ONS report, if not from April 2020.

Concerns about testing in commercial laboratories were documented by the ONS as early as May 2020 [15], when the REACT study discovered that circa 40% of positive tests from commercial laboratories were in fact false positives. A similar false positive rate (44%) was reported in Australia [16] in April 2020. More recently Mr Nicholas Lewis claims that, despite very low false positive rates (0.033%) from testing done by non-commercial and academic laboratories, there may be reason to suspect the operational false positive rates from lighthouse laboratories may be worse than these by some orders of magnitude [17].

Obviously, there is a higher risk of encountering false positives when testing for single genes alone, because of the possibility of cross-reactivity with other human coronaviruses (HCOVs) and prevalent bacteria or reagent contamination. The potential for cross reactivity when testing for SARS-COV-2 has already been confirmed by the German Instand laboratory report from April 2020 [18] (note that Prof. Drosten, co-author of Corman et al [2] is a cooperating partner listed in this report). The report describes the systematic blind testing of positive and negative samples anonymously sent to 463 laboratories from 36 countries and evaluated for the presence of a variety of genes associated with SARS-COV-2[6]. They reported significant cross reactivity and resultant false positives for OC43, and HCoV 229E (a common cold virus) as well as for SARS-COV-2 negative samples, not containing any competing pathogen. Likewise, 70 Dutch laboratories were surveyed in November 2020, by the National Institute for Public Health and the Environment [19], with 76 diagnostic workflows reported as using only one target gene to diagnose the presence of SARS-COV-2 (46% of all workflows).

## Conclusions

Without diagnostic validation, for both the original virus and any variants, it is not clear what can be concluded from a positive test resulting from a single target gene call, especially if there was no confirmatory testing. Many of the reported positive results may be inconclusive, negative or from people who suffered past infection for SARS-COV-2. Even with diagnostic validation of the single target gene call, the UK lighthouse laboratories appear not to be in strict conformance with the WHO emergency use assessment and the manufacturer instructions for use. Given this it is clear the ONS and the UK lighthouse laboratories needs to publicly clarify their use of, and justify the reasons for, deviating from these standards.

[6] N, E, S, ORF1a, ORF1ab and RdRP